\documentclass[twocolumn,showpacs,preprintnumbers,amsmath,amssymb,superscriptaddress,prb]{revtex4}

\usepackage{graphicx}% Include figure files
\usepackage{dcolumn}% Align table columns on decimal point
\usepackage{bm}% bold math

\begin{document}

\title[Noncollinear magnetism and half-metallicity in biased BZGNRs]{Noncollinear magnetism and half-metallicity in biased bilayer zigzag graphene nanoribbons}
\author{Lihua Pan}
\affiliation{School of Physics Science and Technology, Yangzhou University, Yangzhou 225002, China}

\author{Jin An}
 \email{anjin@nju.edu.cn}
\affiliation{National Laboratory of Solid State Microstructures and
Department of Physics, Nanjing University, Nanjing 210093, China}

\author{Yong-Jun Liu}
\affiliation{School of Physics Science and Technology, Yangzhou University, Yangzhou 225002, China}

\begin{abstract}
We study in this paper the edge polarizations and their consequences for a biased Bernal stacked bilayer graphene nanoribbon with zigzag termination. The magnetic states are classified according to the interlayer and intralayer couplings between the edge polarizations, and the magnetic phase diagram of doping versus bias voltage is given. Coplanar magnetic phase is found and the variation of its magnetic structure with the bias voltage is investigated. For all the magnetic states, we also discuss the possibility of the half-metallicity, and for a ribbon with perfect zigzag edges we predict seven kinds of the half-metallic states, which are characterized by their distinct magnetic structures and quantized electrical conductances along the ribbon.
\end{abstract}
\date{\today}% It is always \today, today,
             %  but any date may be explicitly specified
\maketitle
\section{Introduction}
Bilayer graphene has attracted considerable attention due to its unusual electronic properties and prospective applications in graphene-based spintronics\cite{CastroNeto2009,Oleg2010}. While the low-energy excitations in monolayer graphene are massless Dirac fermions with linear dispersion and carry Berry phase $\pi$\cite{ZhouSY2006}, the chiral quasiparticles in Bernal stacked bilayer graphene show finite mass with parabolic dispersion and carry a $2\pi$ Berry phase\cite{Novoselov2006}. For bilayer graphene at charge neutral point, different to the gapless feature of single-particle excitations, theories\cite{HKMin10,FZhang10,JJung11,FZhang11,FZhang12} recently predict some broken symmetry states due to strong electron-electron interactions, which are characterized by small finite gaps, supported by some recent experiments\cite{BEFeldman09,RTW10,FFreitag12,JVelasco12,WBao2012}.

Another strategy for achieving an interaction-induced gap for bilayer graphene is to construct a nanoribbon, especially with zigzag termination\cite{EVCastro2008, LJiao2009}, where the gap opening is due to the interedge interactions between the spin polarizations of the localized edge states. The edge magnetism of zigzag-terminated graphene nanoribbon is of great interest for many scientists even though strong experimental evidence for its existence is still absent\cite{Kunstmann2009,BHuang2008}. According to the Stoner criterion, at or near half-filing, the ground state has a ferromagnetic (FM) instability due to electron-electron interactions and will then lead to an antiferromagnetic (AF) structure where each edge is ferromagnetically polarized but the edges are coupled with each other antiferromagnetically\cite{Fujita1996,Fujita1996B}. In the low carrier doping regime, it was predicted that besides the collinear magnetic states, the monolayer zigzag graphene nanoribbon(MZGNR) can have a noncollinear(NC) magnetic ground state\cite{Jung2009a,Sawada2009,Jung2010}.

Bilayer zigzag graphene nanoribbons(BZGNRs) under a bias voltage provide an additional way to tune the electronic properties of graphene-based spintronics\cite{JianLi2011,EVCastro2011,Padilha2011,Lemonik2012,Guclu2011}, since charge transfer between the layers affects dramatically the bulk and edge states. Thus the bilayer-graphene-ribbon-based electronic devices with the magnetic and transport properties tuned by an electric filed can be expected\cite{TOhta2006,EMcCann2006,Castro2007,PGava2009,JBOostinga2008,ANPal2009,KFMak2009,Zhang2009}. To the best of our knowledge, previous works on the edge magnetism in BZGNR mainly focused on the undoped half-filling case and only the collinear spin polarizations were taken into account\cite{BSahu2008,BSahu2010,MPLima09,YZhang11,YGuo2010,NKarche2011}. Therefore the possibility of stable NC magnetic states in a doped BZGNR becomes a natural question which deserves further investigations.

On the other hand, under the application of a transverse field across a nanoribbon, transition between the semiconducting and half-metallic(HM) states can be achieved by tuning the field strength in both MZGNR\cite{Son2006,Kan2007} and BZGNR \cite{YGuo2010}. For the MZGNR, in the absence of the inversion symmetry, which may be caused by the substrate, it is found that the HM states can be stabilized at both the charge neutrality point\cite{Anjin2011,Soriano2012} and finite doping\cite{Anjin2011}. In analogy with MZGNR, half-metallicity can also be expected in a biased BZGNR due to the existence of edge states and their edge polarizations\cite{GKim2010}.

In this paper, based on the tight-binding model for a biased BZGNR and taking the Coulomb interactions into account, we theoretically explore and classify all the possible magnetic states due to the couplings between edge polarizations and then investigate the possibility of the NC mangnetic states. We also search the possibility of the half-metallicity for all the magnetic phases. It is found that the NC states do exist and can be stabilized as metallic states in a lightly-doping and small-bias regime. Several kinds of HM states are found, which are distinguished by their unique magnetic structures and quantized electric conductance along the ribbon.

The paper is organized as follows. In section 2 we give the model Hamiltonian and its mean-field treatment. In section 3 we classify and discuss the magnetic structures of the ground states and then give the phase diagram. In section 4 we investigate properties of the NC spin canted states and find out all the possible HM states. In Section 5, we summarize our results.

\section{Model}

Different BZGNRs are distinguished by the interlayer stacking arrangements and the edge alignments. Here we study the Bernal stacked BZGNR with the $\beta$ alignment\cite{BSahu2008,BSahu2010}. It consists of two coupled graphene ribbon sheets, where the sublattice A sites of the upper layer are located exactly on the top of sublattice $B$ sites of the lower layer. We assume that the system is subjected to an external electric field between the layers. The model Hamiltonian can be given as follows,
\begin{eqnarray}
\nonumber H=-t\underset{<i,j>\tau\sigma}{\sum}(c_{\tau,i\sigma}^{\dag} c_{\tau,j\sigma}+h.c.)\\
\nonumber -t_{\perp} \underset{<i,j>\sigma}{\sum} \{c_{+,i\sigma}^{\dag} c_{-,j\sigma}+h.c.\}\\
+U \underset{\tau,i}{\sum}(n_{\tau,i\uparrow}-\frac{1}{2})(n_{\tau,i\downarrow}-\frac{1}{2})+
V \underset{i\tau\sigma}{\sum} \tau n_{\tau,i\sigma}
\end{eqnarray}
where $t$ is the intralayer nearest-neighbor hopping integral, and $c_{\tau,i\sigma}$ is the electron annihilation operator at site $i$ with $\sigma=\uparrow, \downarrow$ and $\tau=\pm$ denoting the upper and lower layers. In the second term, the interlayer coupling $t_{\perp}$ is the hopping integral between the $B$ sites on the lower layer and their corresponding top $A$ sites on the top layer. $U$ is the on-site Coulomb repulsion energy, and $2V$ the electrostatic bias voltage. Despite the fact that $V$ term breaks the inversion symmetry, the Hamiltonian at half-filling is still kept invariant under a combination operation of \emph{PI}, where \emph{P} is the particle-hole transformation( $\mathrm{c_{\tau,i\sigma}}\mathrm{\rightarrow}\mathrm{\eta}\mathrm{c^\dag_{\tau,i\sigma}}$, with $\mathrm{\eta=1(-1)}$ if $\mathrm{i}$ belongs to A(B) sublattice of the upper or lower layers) and \emph{I} the space inversion, with the inversion center chosen as the center of the bilayer zigzag ribbon.

Now that the possibility of the noncollinear spin polarization can not be excluded, the Hubbard term is so decoupled that the mean-field Hamiltonian can be written as,

\begin{eqnarray}
\nonumber \mathcal{H}=-t \underset{<i,j>\tau}{\sum}(c_{\tau,i}^\dag c_{\tau,j}+h.c.)+\underset{i\tau}{\sum}(V\tau)n_{\tau,i}\\
\nonumber -t_{\perp}\underset{<i,j>}{\sum}\{c_{+,i}^\dag c_{-,j}+h.c.\}\\
+U\underset{i\tau}{\sum}c_{\tau,i}^\dag(\frac{n_{\tau,i}}{2}
-\mathbf{m}_{\tau,i}\cdot\bm{\sigma})c_{\tau,i}+E_{0},
\end{eqnarray}
where the electron spin polarization and charge density are given by $\mathbf{m}_{\tau,i}=(1/2)\langle c_{\tau,i}^\dag \bm{\sigma}c_{\tau,i}\rangle$, and $n_{\tau,i}=\langle c_{\tau,i}^\dag c_{\tau,i}\rangle$, respectively, with $\bm{\sigma}=(\sigma_{x},\sigma_{y},\sigma_{z})$ the spin Pauli matrices and $c_{\tau,i}^\dag=(c_{\tau,i\uparrow}^\dag, c_{\tau,i\downarrow}^\dag)$. The last constant term is $E_{0}=-U\sum_{i\tau}(n_{\tau,i}^2/4-\mathbf{m}_{\tau,i}^2)$. We choose $t\approx2.8eV$, $U/t=1.0$ and $t_{\perp}/t=0.1$, which are the established values for the coupling parameters\cite{Brandt1988,Dresselhaus2002,Pisani2007,Castro2008}. For this intermediate $U$, the decoupling process introduced is believed to be reliable and effective.

We consider a strip bilayer graphene sample with zigzag termination which is periodic along the vertical direction but open along transverse direction. Equation (2) can be diagonalized to be $\mathcal{H}=\sum_{n} E_{n}\gamma_{n}^{\dagger}\gamma_{n}+E_{0}$ by introducing quasiparticle annihilation operators $\gamma_{n}$ via $c_{\tau,i\sigma}=\sum_{n} u_{\tau,i\sigma}^{n}\gamma_{n}$, with $u_{\tau,i\sigma}^{n}$ the expansion coefficients or amplitudes of the eigenstate corresponding to eigenvalue $E_{n}$. At zero temperature, the self-consistent parameters are given by
$n_{\tau,i}=\sum_{n=1}^{N_{e}}\sum_{\sigma}|u_{\tau,i\sigma}^{n}|^{2}$ and $\mathbf{m}_{\tau,i}=\sum_{n=1}^{N_{e}}(1/2)\sum_{\alpha\beta}u_{\tau,i\alpha}^{n*}\bm{\sigma}_{\alpha\beta}u_{\tau,i\beta}^{n}$. Here, to get doping dependent properties, we fix the total electron number $N_{e}$ rather than the chemical potential of the system. Starting from different initial random spin and density configurations, several self-consistent magnetic solutions may be found by the standard iteration method. By comparing their total energies $(\sum_{n=1}^{N_{e}}E_{n})+E_{0}$ among these solutions, the ground state can be determined by finding out the lowest-energy one. In the following discussion, the width of a BZGNR $W$ is defined as the number of zigzag chains per layer, while the effective doping value is defined as $\delta n=(1-n)\times 2W$ with $n$ the average electron number per site.

\begin{figure}
\hspace{1.0in}
\includegraphics[width=0.5\textwidth,bb=12 10 365 275]{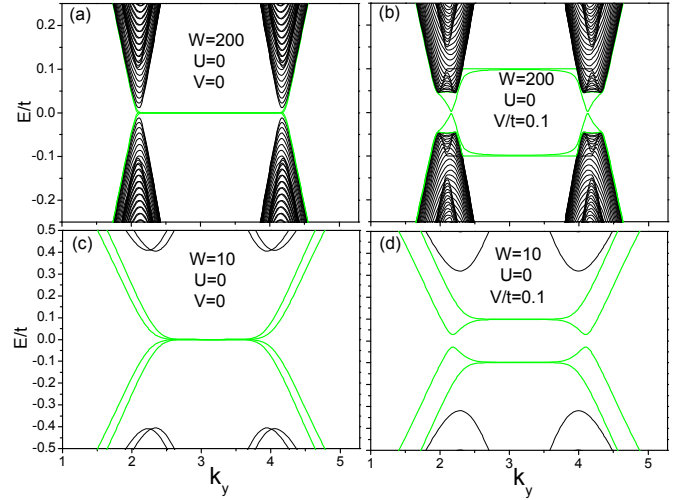}
\caption{\label{Fig. 1}(Color online) Band structure for a BZGNR with width $W$ in the absence of Coulomb interactions($U=0$): (a)$V/t=0$, $W=200$; (b)$V/t=0.1$, $W=200$; (c)$V/t=0$, $W=10$; (d)$V/t=0.1$, $W=10$. $2V$ is the electrostatic bias voltage. The width of a BZGNR $W$ is defined as the number of zigzag chains per layer and energy is measured in unit of $t=2.8eV$.}
\end{figure}

\section{The collinear magnetic structures and the phase diagram}
\begin{figure}
\hspace{1.0in}
\includegraphics[width=0.5\textwidth,bb=0 0 398 657]{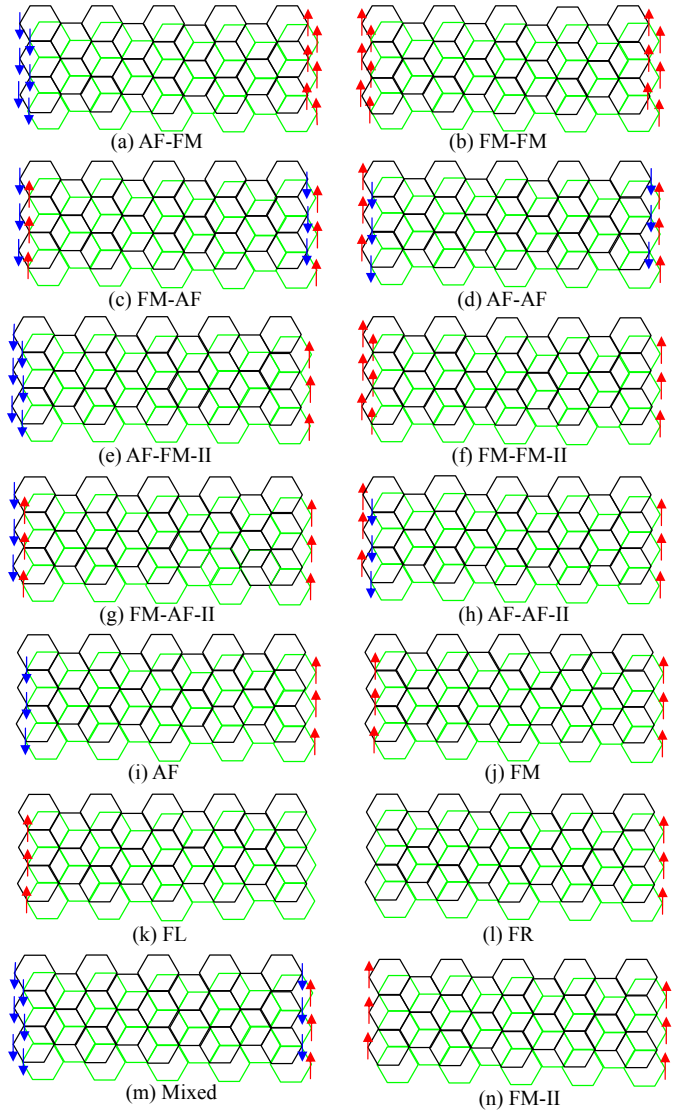}
\caption{\label{Fig. 2}(Color online) Fourteen types of collinear magnetic structures for a BZGNR which has a width $W=10$ and is periodic along the vertical direction. The black(green) hexagons denote the upper(lower) layer lattices while the red(blue) arrows represent the spin-up(spin-down) polarizations at the lattice sites one the edges.}
\end{figure}

In the absence of the Coulomb interactions between electrons, an unbiased BZGNR has four partly flat bands at $E=0$ which correspond to the four edge states with two per edge, while for a biased BZGNR, the four degenerate flat bands are split into two families, each of which consists of one flat edge band and one nearly flat edge band\cite{EVCastro2008}. The band structures in these two situations are demonstrated in figure ~\ref{Fig. 1}. Therefore, at or near half-filling, according to Stoner criteria for magnetic instability, when the Coulomb interactions are taken into account, no matter how small they are, edge polarizations are expected to occur\cite{Castro2008,Fujita1996,MPLima2010}. For a BZGNR with a width $W=10$, by an extensive numerical investigation on model Hamiltonian (2), we find many magnetic states with edge spin polarizations. These magnetic states consist of fourteen types of collinear configurations(figure~\ref{Fig. 2}), as well as the NC canted states including both the symmetric and asymmetric ones. In this section we focus on the collinear states and leave the discussion on the NC ones to next section. Similar to MZGNR, one can classify the fourteen collinear states into five categories, according to the intralayer and interlayer couplings between the edge polarizations.

\begin{figure}
\hspace{0.6in}
\includegraphics[width=0.5\textwidth,bb=0 0 462 327]{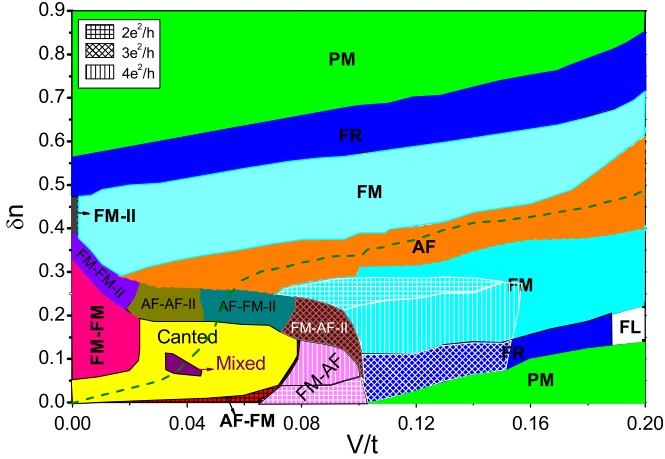}
\caption{\label{Fig. 3}(Color online) The magnetic phase diagram of a BZGNR with width $W=10$: doping value $\delta n$ versus the bias voltage $V/t$, where the effective doping value $\delta n=(1- n)\times 2W$ is measured as the density departure from half-filling per unit cell of the bilayer ribbon, with $n$ the average electron number per site. The dashed line(olive) marks the states in which the lower layer is at half-filling. The HM regimes are denoted and distinguished by the grid, crossing and vertical dashed lines, which represent that the corresponding HM states possess distinct values of the quantized electric conductance $2e^{2}/h$, $3e^{2}/h$ and $4e^{2}/h$ respectively, along a perfect ribbon.}
\end{figure}

In figure~\ref{Fig. 3}, the magnetic phase diagram for a biased BZGNR with width $W=10$ is given. In the light-doping and small-bias regime, both edges are spin polarized. When both the edge spin polarizations are FM(AF), there are two kinds of collinear configurations(see Fig.~\ref{Fig. 2}(a)-(d)), which are denoted by AF-FM or FM-FM(AF-AF or FM-AF), according to whether the coupling between the left and right edges on the same layers is AF or FM. Particularly, when both edges are spin AF polarized, the interedge coupling is expected to be rather small for a relatively wide ribbon, leading to the nearly degenerate AF-AF and FM-AF states\cite{BXu2011}. There also exists the possibility that the edges has different spin polarizations. The state denoted by figure~\ref{Fig. 2}(m) is called "Mixed", since it is spin FM polarized at the left edge but AF polarized at the right one. These five states fall into the first category of the magnetic states. Upon further doping at finite bias, the right edge of the upper layer firstly becomes depolarized with the other spin polarizations qualitatively unchanged. These states are labeled by us by following the denotations of the states of the first category a symbol "-II". Correspondingly four states(shown in figure~\ref{Fig. 2}(e)-(h)) are found to be stabilized, and are classified as the second category. Further doping and increasing $V$ will further depolarize the system. The third category consists of all the possible states with the upper layer fully spin depolarized. Since in this category, only the lower layer are spin polarized, analogously to MZGNR case, one can denote simply the magnetic states as AF or FM(see figure~\ref{Fig. 2}(i),(j)), according to the nature of their interedge coupling. The fourth category includes two states where only one edge of the lower layer is spin polarized, as shown in figure~\ref{Fig. 2}(k)-(l). These two states are denoted by FR and FL respectively. The fifth category consists of only one state(see figure~\ref{Fig. 2}(n)), which lies near $V=0$ at finite doping and is denoted by FM-II, possessing the FM coupled edge polarizations located at the left edge of the upper layer and right edge of the lower layer.

\begin{figure}
\hspace{1.0in}
\includegraphics[width=0.5\textwidth,bb=9 4 401 157]{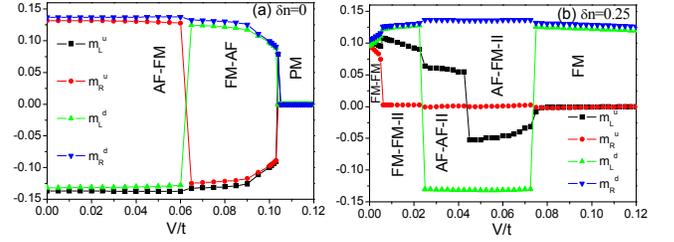}
\caption{\label{Fig. 4}(Color online)Amplitudes of the four edge polarizations as functions of the bias $V/t$ (a) at half-filling, and (b) at $\delta n=0.25$. Superscripts and subscripts in $m_{L(R)}^{u(d)}$ denote edge polarizations of the edge sites of the up(down) layer and at the left(right) edge, respectively.}
\end{figure}

At half-filling, these exist two magnetic phases where the AF-FM phase is stabilized at lower $V$ while the FM-AF one is stabilized instead at larger $V$. When $V$ is larger than the critical value $V_{c}\approx0.105t$, the BZGNR system will undergo a phase transition from the FM-AF state to a paramagnetic(PM) state. As mentioned before, the model Hamiltonian at half-filling is invariant under \emph{PI}. It is found that all the above three states at half-filling preserve the \emph{PI} symmetry and so each of them has the following symmetry properties:
\begin{eqnarray}
\nonumber n_{L}^{u}+n_{R}^{d}=2&,&\ \ \ n_{L}^{d}+n_{R}^{u}=2\\
\mathbf{m}_{L}^{u}=-\mathbf{m}_{R}^{d}&,&\ \ \  \mathbf{m}_{L}^{d}=-\mathbf{m}_{R}^{u}.
\end{eqnarray}
where $n_{L(R)}^{u(d)}$ and $\mathbf{m}_{L(R)}^{u(d)}$ are the electron densities and edge spin polarizations at the sites on the four edges of the two layers. This \emph{PI} symmetry is even preserved near the transition point between the AF-FM and FM-AF states, while it is completely lost when the system is away from half-filling (See figure~\ref{Fig. 4}).

On the other hand, in the absence of a gate bias, i.e., at $V=0$, the model system is inversion symmetric, which means that the ground state might be invariant under inversion \emph{I}. We find that upon increasing $\delta n$, the inversion symmetric and inversion asymmetric states occur alternatively. Concretely, the ground state is found to be inversion symmetric for all the FM-FM states($0.05<\delta n\leq0.325$) and all the FM-II states($0.38<\delta n<0.475$) at $V=0$, while found to be inversion asymmetric otherwise. There exist a regime of symmetric NC canted states located at $0<\delta n\leq0.02$, which is $\sigma_{v}$\emph{I} symmetric apart from a global spin rotation, with $\sigma_{v}$ the mirror reflection with respect to the plane normal to the zigzag direction. All the above inversion symmetric collinear states satisfy:

\begin{eqnarray}
\nonumber n_{L}^{u}=n_{R}^{d}&,& \ \ \ n_{L}^{d}=n_{R}^{u} \ \ \\
\mathbf{m}_{L}^{u}=\mathbf{m}_{R}^{d}&,& \ \ \  \mathbf{m}_{L}^{d}=\mathbf{m}_{R}^{u},
\end{eqnarray}
while the symmetric coplanar NC states satisfy:
\begin{eqnarray}
\nonumber n_{L}^{u}=n_{R}^{d},\ \ \ n_{L}^{d}=n_{R}^{u} \ \ \ \ \ \ \ \ \ \ \ \ \ \ \ \ \ \ \ \ \\
|\mathbf{m}_{L}^{u}|=|\mathbf{m}_{R}^{d}|, |\mathbf{m}_{L}^{d}|=|\mathbf{m}_{R}^{u}|,
\mathbf{m}_{L}^{u}\times\mathbf{m}_{L}^{d}=\mathbf{m}_{R}^{u}\times\mathbf{m}_{R}^{d}
\end{eqnarray}
where the relative angle between $\mathbf{m}_{L}^{u}$ and $\mathbf{m}_{L}^{d}$ varies from $0$ to $33^{\circ}$.

Once $V>V_{c}$, sufficient electrons will be transferred from the upper to the lower layer, making the upper layer a PM state without polarizations at both edges, regardless of the doping level. Since the upper layer is fully depolarized, the magnetism of the BZGNR is dominated by the lower layer and then the system can be viewed as one possessing approximate particle-hole symmetry at half-filling of the lower layer. As a consequence, the magnetic states together with the PM ones are found to distribute nearly symmetrically about the half-filling line of the low layer(the dashed line in figure~\ref{Fig. 3}) in the phase diagram. The other consequence is that starting from an AF state, upon doping, the system undergoes transitions one by one to the FM, to the FR, and to the PM states respectively, analogous to that in MZGNR\cite{Jung2009a,Anjin2011}. Furthermore, except for the NC canted states, all phases located at this dashed line, including the AF, the AF-FM-II and the Mixed states, have AF couplings between the left and right edges on the lower layer, which is consistent with that in MZGNR\cite{Fujita1996,Pisani2007,HLee2005}.

The topological structure of the phase diagram is nearly unchanged for different ribbon width $W$ as long as it is less than 20. For a wider ribbon, we note that since the interedge coupling is more negligible, especially for the case where one of the edges is spin AF polarized\cite{BXu2011}, the FM-AF and AF-AF states(or the FM-AF-II and AF-AF-II states) become competing and in certain regime of phase diagram, they are nearly degenerate.

\section{The canted states and the HM states}

\begin{figure}
\hspace{1.0in}
\includegraphics[width=0.5\textwidth,bb=0 0 406 555]{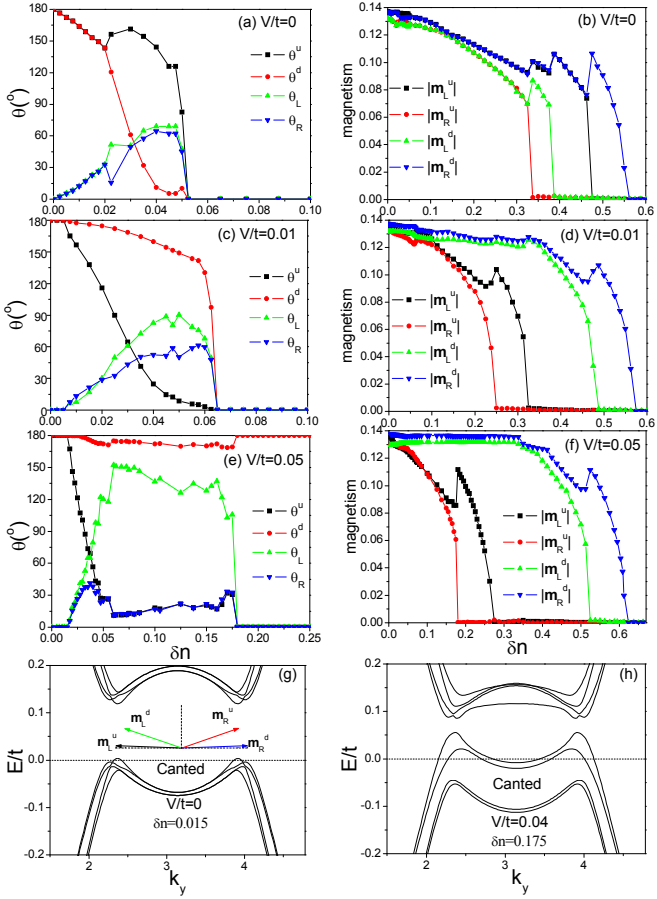}
\caption{\label{Fig. 5}(Color online)The left panels (a), (c) and (e) exhibit the relative orientation angles between the spin polarization directions, while the right panels (b), (d) and (f) exhibit the magnitudes of the four edge polarizations as functions of doping, at different biased voltages $V$. Two representative band structures of the NC canted states are shown in (g) at $V/t=0$, $\delta n=0.015$ for a symmetric one, and in (h) at $V/t=0.04$, $\delta n=0.175$ for an asymmetric one. $\theta^{u}$($\theta^{d}$) are the relative angles between the left and right edge spins on the upper(lower) layer, and $\theta_L$($\theta_R$) are the relative angles between the left(right) edge spins on the upper ad lower layers. $|\mathbf{m}_{L(R)}^{u(d)}|$ are defined the same as that in figure~\ref{Fig. 4}.}
\end{figure}

Now let's focus on the magnetic properties of the NC canted states. It can be seen from the phase diagram that the NC canted states exist in the low-doping and small-bias regime, which is much larger, compared with that in MZGNR\cite{Jung2009a,Anjin2011}. So the magnetic interlayer interactions really alter the character of the magnetic states. All the NC canted states are found to be coplanar and so can be characterized by the magnitudes of and the relative orientation angles between edge polarizations. We denote $\theta^{u}$($\theta^{d}$) as the relative angles between the left and right edge spins on the upper(lower) layer, and $\theta_L$($\theta_R$) as the relative angles between the left(right) edge spins on the upper ad lower layers.

These NC canted states are always found to be metallic, as can be seen by the band fillings of the two typical states shown in figure~\ref{Fig. 5}(g)and (h). It can be seen that among the eight split edge bands due to the bias voltage as well as the Coulomb interactions, one(two) band(s) is(are) partly filled for the first(second) case. Therefore, dependent on doping, the spectrum is characterized by four or eight counterpropagating current-carrying states at the Fermi level, leading to a quantized electric conductance $2e^{2}/h$ or $4e^{2}/h$ along the ribbon with perfect edge, respectively.

\begin{figure}
\hspace{1.0in}
\includegraphics[width=0.5\textwidth,bb=9 2 372 529]{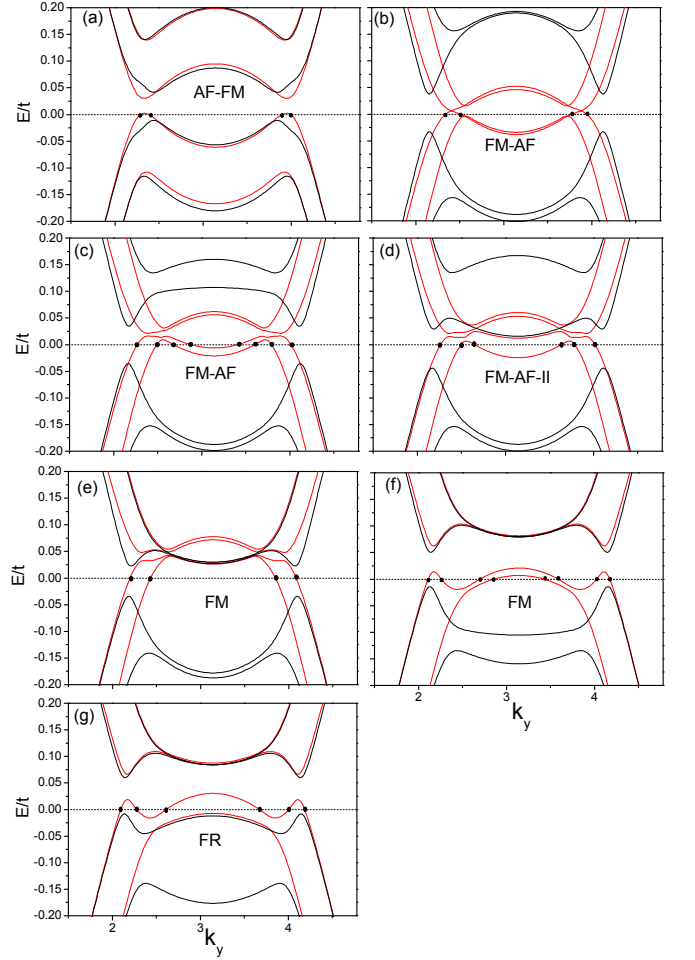}
\caption{\label{Fig. 6}(Color online) Band structures for the seven typical HM states for the BZGNR with width $W=10$: (a) $V/t=0.06$ and $\delta n=0.015$; (b) $V/t=0.09$ and $\delta n=0.025$; (c) $V/t=0.09$ and $\delta n=0.125$; (d) $V/t=0.09$ and  $\delta n=0.1625$; (e) $V/t=0.09$ and $\delta n=0.2625$; (f) $V/t=0.14$ and $\delta n=0.1375$; (g) $V/t=0.14$ and $\delta n=0.1125$.}
\end{figure}

Three representative doping processes related to the NC canted states are shown in figure~\ref{Fig. 5}(a),(c),(e). For the unbiased BZGNR $V=0$, the BZGNR at half-filling is the AF-FM insulating state\cite{YZhang11}. Upon doping the ground state undergoes a continuous phase transition to a $\sigma_{v}$\emph{I} symmetric NC canted state, which is characterized by the relations $\theta^{u}=\theta^{d}$, $\theta_{L}=\theta_{R}$. When $0.02<\delta n< 0.052$, the NC ground states becomes asymmetric, i.e., $\theta^{u}\neq \theta^{d}$ and $\theta_{L}\neq \theta_{R}$.

figure~\ref{Fig. 5}(b),(d),(f) give the edge magnetizations of a BZGNR as functions of doping for different $V$, which clearly exhibit the depolarization process upon doping. In particular, the latter two show the successive depolarization processes via the first-order transitions for cases at finite bias. The right edge of the upper layer becomes firstly depolarized, then the left edge of the upper layer, the left edge of the lower layer, and finally the right edge of the lower layer. This is interpreted as follows. When a BZGNR is doped away from half-filling, a finite gate bias will transfer electrons from the upper to the lower layer and then drive the upper layer further away from half-filling while drive the lower layer closer to half-filling, making this layer easier to be spin-polarized. On the other hand, the left(right) edge sites of the upper(lower) layer are most easily spin polarized since they are actually the leftmost(rightmost) sites of the BZGNR, which are closer to magnetic instability.

In analogy with MZGNR, the half-metallicity can be expected in BZGNR. The regimes where the HM states exist are clearly demonstrated in the phase diagram. Different HM states can be distinguished by their magnetic orders and the values of their quantized electrical conductances along the ribbon, the latter of which can be determined by counting the number of the current-carrying states at the Fermi level, i.e., the number of the crossing points between the band spectrum and the chemical potential. From the phase diagram, we see that all the HM states form a connected regime below the half-filling line of the lower layer, indicating that the HM states only exist where the doping level of the lower layer is more than half-filling. We find seven kinds of the HM states in five magnetic phases(see figure~\ref{Fig. 6}): The AF-FM phase at finite doping is HM with a quantized electric conductance $2e^{2}/h$; The FM-AF phase is HM and has a quantized electric conductance $2e^{2}/h$ at low doping but $4e^{2}/h$ at larger doping. The FM-AF-II phase are HM one characterized by a quantized electric conductance $3e^{2}/h$. Two parts of the FM phase are HM with quantized electric conductance $2e^{2}/h$ and $4e^{2}/h$ respectively. Finally, one part of the FR phase is HM, possessing a quantized electric conductance $3e^{2}/h$. Thus for a BZGNR at a fixed doping, one can achieve phase transitions between different HM states by tuning the bias voltage, which may have great applications in graphene manipulation.

\section{Summary}

In summary, we have investigated the possibility of the noncollinear spin canted states and half-metallicity in a bilayer zigzag nanoribbon under a bias voltage. Due to the interlayer and intralayer couplings between the edge polarizations, the magnetic states are classified and the phase diagram is given. At a small but finite bias voltage, upon doping away from half-filling, the bilayer ribbon will be spin depolarized step by step: firstly the right-edge of the upper layer, then the left-edge of the upper layer, the left-edge of the lower layer, and finally the right-edge of the lower layer. Both symmetric and asymmetric noncollinear spin canted states are found to be stabilized as metallic ones in the low-doping and small-bias regime. Seven kinds of the HM states are found, each of which has distinct magnetic structure or different quantized electrical conductance along the perfect ribbon. Since the HM properties are important in spin-related transport, these results are expected to have potential applications in bilayer-graphene-based spintronics.

\begin{acknowledgments}
This work was supported by NSFC Projects No.11174126, and 973 Projects No. 2011CB922101.
\end{acknowledgments}

\end{document}